\documentstyle[prd,aps,floats,tabularx,epsfig]{revtex}
 \newcommand{\be}{\begin{equation}} \newcommand{\ee}{\end{equation}} 
\newcommand{\bea}{\begin{eqnarray}} 
\newcommand{\eea}{\end{eqnarray}}

\newcommand {\simlt}{\lower.5ex\hbox{$\; \buildrel < \over \sim \;$}} 
\newcommand {\simgt}{\lower.5ex\hbox{$\; \buildrel > \over \sim \;$}}

\begin{document} 
\setlength{\unitlength}{1mm} 
\twocolumn[\hsize\textwidth\columnwidth\hsize\csname@twocolumnfalse\endcsname 
\title{On the Density of Cold Dark Matter.} 
\author{Alessandro Melchiorri$^{\sharp}$ and Joseph Silk$^{\sharp}$} 
\address{$^{\sharp}${\it Denys Wilkinson Building, University of Oxford,  
Keble Road, Oxford, OX 3RH, UK.}} 
\maketitle 
\begin{abstract} 
 
 The nature of dark matter 
is increasingly constrained by cosmological data. 
In this paper, 
we examine  the implications of the Cosmic Microwave  
Background anisotropy limits on  the  
density  of cold dark matter  under  
 different theoretical assumptions and  
combinations of  datasets. We infer the constraint  
$\Omega_{cdm}h^2=0.12\pm0.04$ (at $95 \%$ c.l.).
 The CDM models are compared with the shape of the linear  
matter power spectrum inferred from the 2dF galaxy redshift survey and  
with the rms mass fluctuations from recent local cluster observations. 
We found that a value of $\sigma_8 \sim 1$ as  suggested by
recent cosmic shear data is not favoured by the CMB data alone 
nor  by combined CMB+SN-Ia, CMB+HST or CMB+2dFGRS analyses. 
We also extrapolate our bounds on the rms linear mass fluctuations 
to sub-galactic scales and compare them with recent 
lensing constraints, finding agreement with the standard 
$\Lambda$CDM model. 
\end{abstract} 
\bigskip] 
 
\section{Introduction} 
 
The new results on Cosmic Microwave Background (CMB) anisotropy 
from TOCO (\cite{toco}, BOOMERanG~(\cite{boom97}, \cite{boom98}),  
MAXIMA~(\cite{maxima}), and DASI~\cite{Dasi} experiments represent 
an extraordinary success for the 
standard cosmological model of structure formation based on 
Cold Dark Matter (CDM) and adiabatic primordial perturbations  
(see e.g. \cite{th4b}).  
Furthermore, early data releases from the 2dFGRS and SDSS  
galaxy redshift surveys (\cite{2dfgrs}, \cite{sdss}) are living 
up to expectations and combined analysis of all these datasets  
are placing strong constraints on most cosmological parameters  
(~\cite{efstathiou},\cite{lahav}). 
 
However, even if theory and observations are in spectacular  
agreement on large ($\simgt 1 \rm Mpc$) scales,  
various discrepancies seem to  be present on 
smaller (sub-galactic) scales. In particular, numerical simulations of  
CDM models yield an excess of small-scale power, producing, for 
example,  more satellite galaxies than observed  (\cite{moore}) and 
cuspy galactic halos 
with excessive dark mass concentrations (\cite{binn}, \cite{sell}), 
while  
on the smallest scales probed, gravitational lensing (\cite{dalal},\cite{metcalf}) prefers CDM substructure.
 Many authors have addressed these issues and many solutions have been proposed, ranging from modifications of 
the properties of the CDM particles to the introduction of new physics 
and to a refinement of the astrophysical processes involved in the 
simulations. 

One way to address these issues is by
detection of dark matter. The key quantity of relevance
to dark matter searches is the CDM density, which specifies the dark matter annihilation cross-section.
In this  paper,
we use the latest CMB and deep redshift surveys  
data in order to  simultaneously constrain the  
most important cosmological observables 
related to CDM: the CDM density
$\Omega_{cdm}h^2$, the spectral tilt $n_S$ of the primordial power spectrum and the  
rms mass fluctuations in spheres of $8h^{-1}\rm Mpc$, $\sigma_8$. 
 
We extend and complement  recent 
studies  (see e.g. \cite{efstathiou}, \cite{lahav}, \cite{thx})
by studying the stability of the constraints 
towards different theoretical assumptions such as, for example, 
a dark energy equation of state parameter $w_{Q}>-1$ (quintessence) 
or an extra background of relativistic particles  
($\Delta N_{\nu}^{eff}>0$). 
 Furthermore, under the assumption of CDM,  
we extrapolate the cosmological constraints  
into a measurement of the linear rms density fluctuations on  
sub-galactic mass scales.  

\section{Cosmological Constraints on CDM} \label{cmb} 
 
\subsection{Method} 
 
\medskip

\begin{figure}[htb] 
\begin{center}
\epsfxsize=6.6cm
\epsfysize=5.2cm
\epsffile{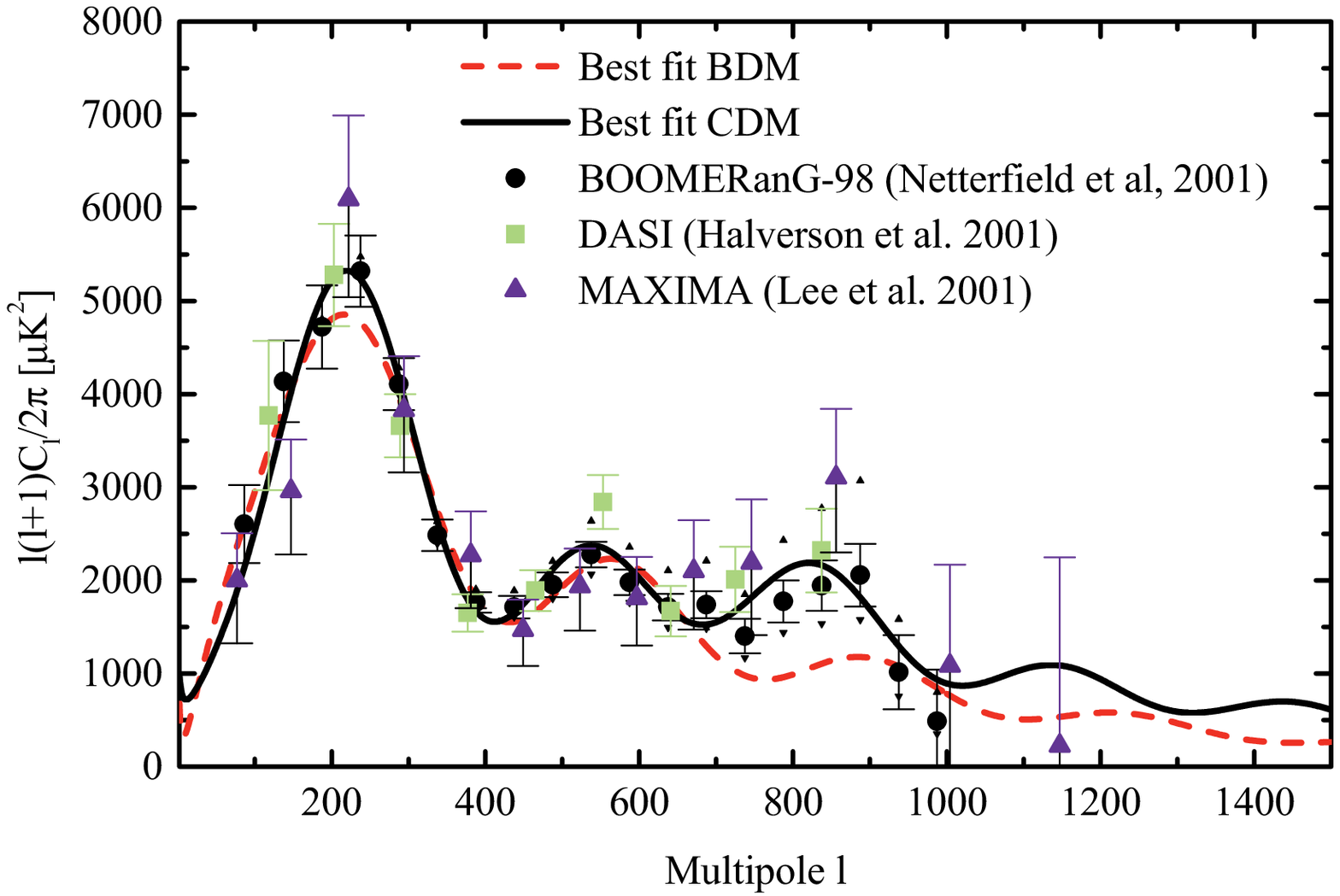}
\epsfxsize=6.5cm
\epsfysize=5.2cm
\epsffile{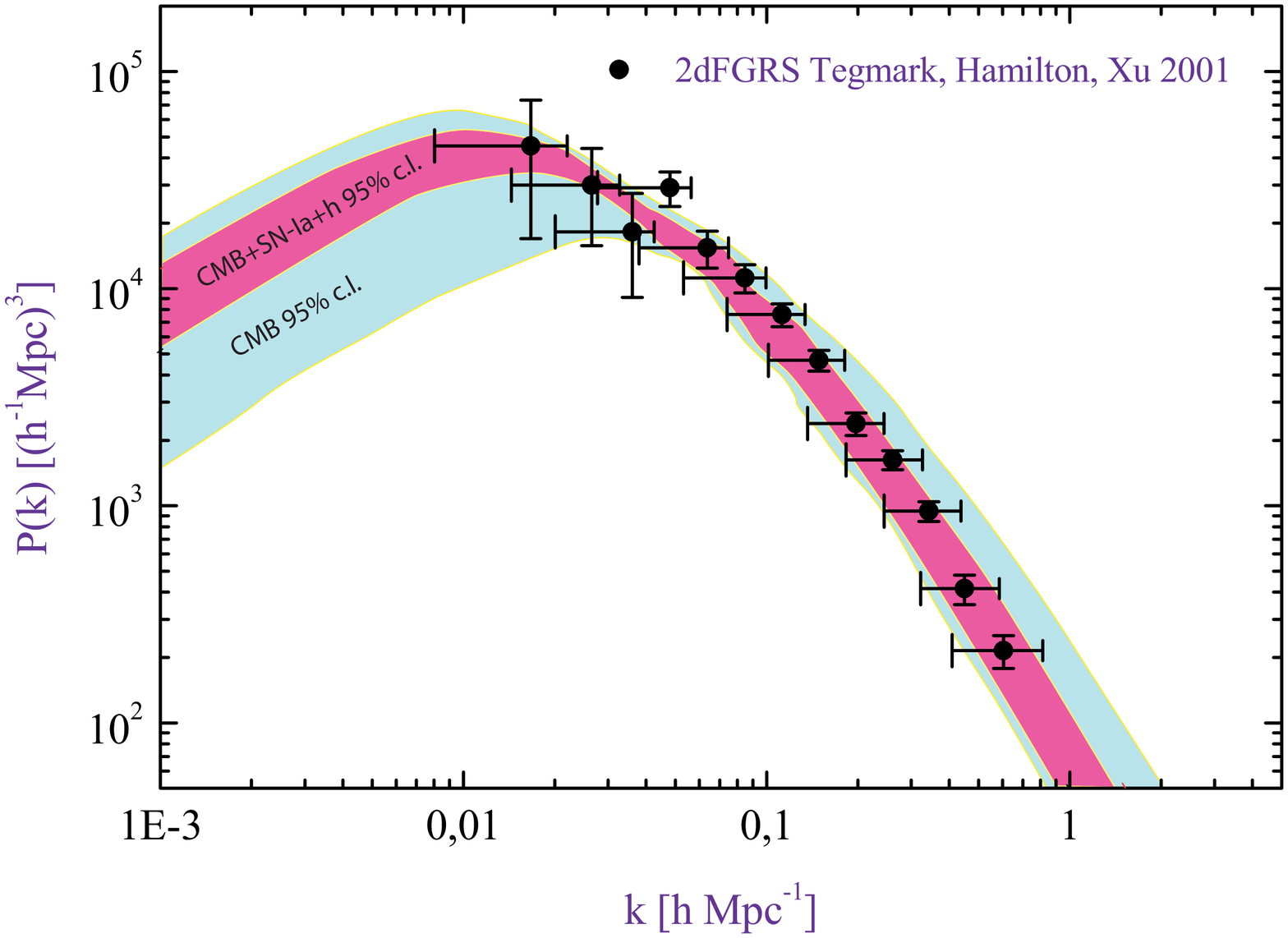}
\end{center}
\caption{Top panel: CMB anisotropies and CDM. Bottom panel:
Allowed region for the
matter power spectrum from CMB and from other cosmological 
observables obtained under the assumption of adiabatic CDM primordial 
fluctuations. The data from the 2dF redshift survey  
(Tegmark and Hamilton, 2002) is also plotted in the figure.} 
\label{fig:matpower} 
\end{figure} 
 
\medskip

\begin{figure}[htb] 
\begin{center} 
\epsfxsize=8.0cm 
\epsfysize=6.5cm 
\epsffile{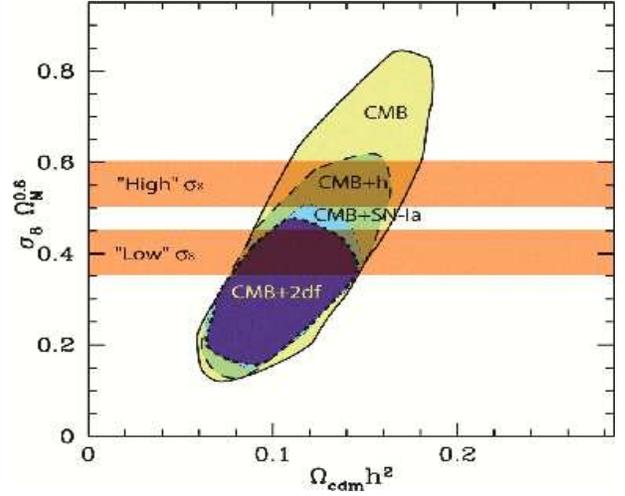} 
\end{center} 
\caption{Constraints in the $(\Omega_m)^{0.6}\sigma_8-\Omega_{cdm}h^2$ 
plane. The results of the $3$ combined analysis CMB+HST, CMB+SN-Ia and 
CMB+2dFGRS are shown together with the $68 \%$ c.l. cluster constraints} 
\label{fig:bmps2.eps} 
\end{figure} 
\medskip 
 
We compare  the recent CMB observations with a set of  
models with 6 parameters sampled as follows:  
$\Omega_{cdm}h^2\equiv \omega_{cdm}= 0.01,...0.40$, in steps of  $0.01$;  
$\Omega_{b}h^2\equiv\omega_{b} = 0.001, ...,0.040$, 
in steps of  $0.001$; $\Omega_{\Lambda}=0.0, ..., 1.0$, 
in steps of  $0.05$ and  
$\Omega_k$ such that $\Omega_m=0.1, ..., 1.0$, in steps of  $0.05$. 
The value of the Hubble constant is not an independent 
parameter, since: 
 \begin{equation} 
h=\sqrt{{\omega_{cdm}+\omega_b} \over {1-\Omega_k-\Omega_{\Lambda}}}. 
\end{equation} 
 We vary the spectral index of the primordial density perturbations 
within the range $n_s=0.60, ..., 1.40$ (in steps of  $0.02$)  
and we rescale the amplitude of the fluctuations by a 
pre-factor $C_{10}$, in units of $C_{10}^{COBE}$, with  
$0.50 < C_{10} < 1.40$. 
In order to test the stability of our results under variation 
of some of the theoretical assumptions, we also consider  variations in the dark energy equation of state parameter  
$w_Q=-1.0,...,-0.5$ as expected in  
quintessence models (see e.g. \cite{copeland}, \cite{balbi}, 
\cite{th4}, \cite{maor}) and  
of an extra background of relativistic particles with  
$\Delta N_{\nu}=0.3,...,9$ in steps of  $0.3$ (see \cite{bowen}
and references therein). 
 
The theoretical models are computed using a modified version of  
the publicly available {\sc cmbfast} code~\cite{sz} and are compared  
with the recent BOOMERanG-98, DASI and MAXIMA-1 results. 
The power spectra from these experiments were estimated in  
$19$, $9$ and $13$ bins respectively, spanning the range 
$25 \le \ell \le 1150$.  
For the DASI and MAXIMA-I experiments we use the publicly available 
correlation matrices and window functions. 
For the BOOMERanG experiment we assign a flat interpolation  
for the spectrum in each bin $\ell(\ell+1)C_{\ell}/2\pi=C_B$,  
we approximate the signal $C_B$ inside 
the bin to be a Gaussian variable and we consider $\sim 10 \%$  
correlations between the various bins. 
The likelihood for a given cosmological model is then 
 defined by  
$-2{\rm ln} {\cal L}=(C_B^{th}-C_B^{ex})M_{BB'}(C_{B'}^{th}-C_{B'}^{ex})$ 
where  $M_{BB'}$ is the Gaussian curvature of the likelihood  
matrix at the peak.  
 We consider $10 \%$, $4 \%$  and $5 \%$ Gaussian distributed  
calibration errors for the BOOMERanG-98 \cite{Boom2}, DASI  
\cite{Dasi} and MAXIMA-1 \cite{Max2} experiments respectively and  
we included the beam uncertainties by the analytical marginalization 
method presented in \cite{sara}. 
We also include the COBE data using Lloyd Knox's RADPack packages. 

In addition to  the CMB data we incorporate the real-space power spectrum  
of galaxies in the 2dF 100k galaxy redshift survey using the 
data and window functions of the analysis of Tegmark et al. (\cite{thx}). 
 
To compute ${\cal L}^{2dF}$, we evaluate $p_i = P(k_i)$,  
where $P(k)$ is the theoretical matter power spectrum  
and $k_i$ are the $49$ k-values of the measurements in \cite{thx}.  
Therefore we have $-2ln{\cal L}^{2dF} = \sum_i [P_i-(Wp)_i]^2/dP_i^2$, 
where $P_i$ and $dP_i$ are the measurements and corresponding error bars 
and $W$ is the reported $27 \times 49$ window matrix. 
We restrict the analysis to a range of scales where the fluctuations 
are assumed to be in the linear regime ($k < 0.2 h^{-1}\rm Mpc$). 
When combining with the CMB data, we marginalize over a bias $b$  
considered to be an additional free parameter. 
 
We attribute a likelihood  
to each value of $\omega_{cdm}$, $n_S$, $\sigma_8$  
 by marginalizing over the {\it nuisance} parameters.  
We then define our $68\%$ ($95 \%$),  
confidence levels to be where the integral of the  
likelihood is $0.16$ ($0.025$) and $0.84$  
($0.975$) of the total value.

\subsection{CMB results and test for theoretical assumptions.} 

\begin{figure}[htb]
\begin{center}
\epsfxsize=7.0cm
\epsfysize=6.0cm
\epsffile{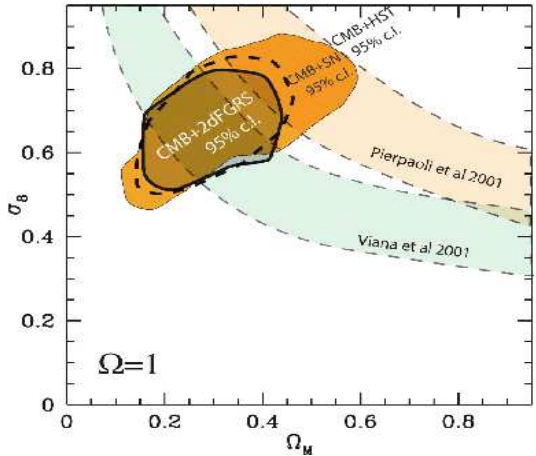}
\end{center}
\caption{Constraints in the $\Omega_m-\sigma_8$ plane.
The results of the $3$ combined analysis CMB+HST, CMB+SN-Ia and
CMB+2dFGRS are shown together with the $68 \%$ c.l. constraints from
Viana et al. 2001 and Pierpaoli et al 2001.}
\label{fig:like_mar2}
\end{figure}
\medskip

\begingroup\squeezetable 
\begin{table}[htb] 
\renewcommand*{\arraystretch}{1} 
\caption{
  Parameter estimates from the CMB 
  data sets and complementary data. 
  Below the line, 
  we restrict the parameter space  
  to $\Omega=1$. $\sigma_8^*$ is the value 
of $\sigma_8$ computed in the region 
 $\Omega_m=0.30\pm0.05$. Central values and $2\sigma$  
limits are found from the 50\%,  
2.5\% and 97.5\% integrals of the marginalized likelihood.} 
\begin{tabular}{lrrrrrrrrrr} 
Case 
& \multicolumn{1}{c}{$\Omega_{cdm}h^2$} 
& \multicolumn{1}{c}{$n_s$} 
& \multicolumn{1}{c}{$\sigma_8^*$} 
\\ 
\hline 
Weak Priors. 
&$0.12^{+0.06}_{-0.06}$   
&$0.92_{-0.10}^{+0.11}$ 
&$0.66_{-0.11}^{+0.11}$ 
\\ 
\hline 
$\Omega=1$ 
& 
& 
& 
\\ 
\hline 
\\ 
$-1.0<  w_Q < -0.5$  
& $0.12^{+0.05}_{-0.05}$ 
& $0.92_{-0.09}^{+0.12}$ 
& $0.63^{+0.13}_{-0.13}$ 
\\ 
$N^{eff}_{\nu}$ free 
& $0.21^{+0.11}_{-0.12}$ 
& $0.92_{-0.09}^{+0.12}$ 
& $0.68^{+0.11}_{-0.11}$ 
\\ 
$\tau_{c}$ free 
& $0.13^{+0.04}_{-0.05}$ 
& $0.94_{-0.09}^{+0.15}$ 
& $0.68^{0.15}_{-0.09}$ 
\\ 
no COBE 
& $0.12^{+0.05}_{-0.05}$ 
& $0.94_{-0.18}^{+0.16}$ 
& $0.63^{+0.12}_{-0.15}$ 
\\ 
$\ell<650$ 
& $0.15^{+0.06}_{-0.05}$ 
& $0.92_{-0.10}^{+0.11}$ 
& $0.72_{-0.11}^{+0.10}$ 
\\ 
SN-Ia 
& $0.11^{+0.03}_{-0.03}$ 
& $0.93_{-0.07}^{+0.06}$ 
& $0.65_{-0.10}^{+0.11}$ 
\\ 
$h=0.71\pm0.08$ 
&$0.12^{+0.04}_{-0.05}$ 
&$0.93_{-0.08}^{+0.08}$ 
&$0.65_{-0.10}^{+0.11}$ 
\\ 
2dF GRS 
& $0.11^{+0.03}_{-0.04}$ 
& $0.93_{-0.05}^{+0.08}$ 
& $0.65_{-0.10}^{+0.10}$ 
\\ 
$\Omega_m^{0.6}\sigma_8=0.5\pm0.05$ 
& $0.14^{+0.03}_{-0.03}$ 
& $0.92_{-0.06}^{+0.06}$ 
& $0.73_{-0.10}^{+0.11}$ 
\\ 
$\Omega_m^{0.6}\sigma_8=0.4\pm0.05$ 
& $0.12^{+0.03}_{-0.03}$ 
& $0.91_{-0.06}^{+0.06}$ 
& $0.68^{+0.10}_{-0.09}$ 
\\ 
\end{tabular} 
\end{table} 
\endgroup 
 
The results of our analysis are shown in Table 1. 
In the first row, we restrict our  
analysis to CMB data with a combination of  
``weak priors'': $h=0.65\pm0.2$, age $t_0>10$Gyr and $\Omega_m>0.1$. 
Since the value of $\sigma_8$ is degenerate with $\Omega_m$, 
in the table we report $\sigma_8^*$ defined as the value 
of $\sigma_8$ with $\Omega_m=0.30\pm0.05$. 
We  see that, under the class of models  considered,  
the CMB data suggest a $\sim 2 \sigma$  
detection of  CDM, a scalar spectral index $n_S\sim 0.92$ 
and a value of $\sigma_8 \sim 0.66$ at $\Omega_m \sim 0.3$. 
The reason why the present data favour CDM is evident in Fig.1,
top panel, where we plot the recent CMB data  
with the best fit purely baryonic (BDM) and CDM models.  
BDM models fail to reproduce the observed power 
at $\ell \ge 700$. The new CMB data provide  independent  
support for the presence of non-baryonic dark matter in the universe.

Before including information from complementary cosmological datasets, 
it is important to test the stability of our CMB results by 
removing  some of the theoretical assumptions used in the  
analysis. 
 We restrict our analysis to $\Lambda$CDM models. However another  
candidate that could possibly explain the observations of an accelerating 
universe is a dynamical scalar ``quintessence'' field.  
The common characteristic of quintessence models is that their equation  
of state,$w_{Q}=p/\rho$, varies with time and can be greater than 
$w_Q=-1$,  the value corresponding to the  cosmological constant.  
Adopting 'quintessence' instead of a cosmological constant does not 
change the results of our analysis, but increases the error bars. 
 
Another possibility is to consider an extra 
background of relativistic 
particles (see e.g. \cite{bowen}), parametrized by a larger 
number of effective massless neutrinos $N_{\nu}^{eff}$. 
Since increasing $N^{eff}_{\nu}$ changes the epoch of equality, 
which is well determined by CMB observations (\cite{bowen}), 
larger values of $\Omega_{cdm} h^2$ are needed in order to compensate 
for this variation. The constraints on  
$\sigma_8^*$ are not greatly affected since the matter power 
spectrum is mainly sensitive to changes in the redshift of equality, 
which is kept constant via the CMB data. 
 
In the entire analysis, we assume  negligible reionization and 
an optical depth $\tau_c\sim0$. This is in agreement 
with recent estimates of the redshift of reionization 
$z_{re}\sim 6 \pm 1$ (see e.g. \cite{gnedin}). 
We have also removed this assumption. Due to the well known  
degeneracy with the scalar spectral index $n_S$, the effect of including 
variations in $\tau_c$  leaves the CMB data  in better agreement 
with higher values of $n_S$ and of $\sigma_8$.
 
 A background of gravity waves and/or of isocurvature modes 
can modify the theoretical CMB spectrum on large angular scales. 
In order to test for these hypotheses, we repeat the analysis  
without the COBE data. This has the effect 
of relaxing our constraints. 
 
The high-$\ell$ part of the observed spectrum can be contaminated  
by different systematics (see e.g. \cite{tegsy}): beam reconstruction, 
detector noise, foregrounds.  
All  $3$ different experiments can be affected 
by different systematics and the fact that the $3$ datasets are in  
reasonable agreement suggests that the systematics  
are under control. However, we repeated the analysis removing the 
data points at $\ell > 650$.  
As we can see, again, apart from an increase in the error bars, 
there is no significant difference. 
 
\subsection{Comparison with complementary datasets.} 
 
Since the CMB results are stable under variations of different 
theoretical assumptions, we can now assume that
the $\Lambda$-CDM models 
are  valid and investigate the effects of applying 
various prior probabilities and of incorporating complementary cosmological 
datasets. 
 Including the gaussian prior $0.8\Omega_m-0.6\Omega_{\Lambda}=-0.2\pm0.1$  
from type Ia supernova luminosity distances \cite{perl}  or 
$h=0.71\pm0.07$ from measurements with the Hubble Space Telescope 
\cite{free} does not change our conclusions and improves the constraints 
on the $3$ parameters. 
 
In Figure $1$, bottom panel, we  check for consistency of the 
2dF data with the set of CDM models used in the CMB analysis by plotting a  
convolution of all the matter power spectra from the theoretical models 
in agreement with the CMB data, together with the recent 
2dF analysis of \cite{thx}. As one can see, the region consistent 
with CMB alone is quite broad (due to the weak CMB constraint 
on $\Omega_{\Lambda}$) and contains the shape of the 2dF spectrum. 
Including other cosmological constraints from SN-Ia and HST shrinks 
the CMB constraint into a region consistent with the shape 
inferred from 2dF. 
 This 'consistency' is reflected in the results of Table 1: 
including information on the shape of the 2dF  
matter power spectrum improves  
our constraints in a similar direction 
to the SN-Ia and HST priors. 
 
On similar scales, recent analyses of the local cluster 
 number counts can be summarised as giving different results for   
$\sigma_8$ mainly due to systematics in the calibration between cluster  
virial mass and temperature:  
a {\it high} value  $\sim \Omega_m^{0.6} \sigma_8=0.55 \pm 
0.05$ in agreement with the results of (\cite{pierpa}, \cite{eke})  
and a {\it lower} one, $\sim \Omega_m^{0.6} \sigma_8=0.40 \pm 0.05$  
following the analyses of \cite{s8eljak} and \cite{liddle}. 
 It is interesting to plot these constraints in the  
$\Omega_m^{0.6}\sigma_8-\Omega_{cdm}h^2$ plane. 
We do this in Figure $2$, where we plot the $95 \%$ confidence level 
contour of the combined CMB+HST, CMB+SN-Ia and CMB+2dFGRS analyses 
(obtained again with the assumption of CDM) together with the  
{\it high} and {\it low} constraints on $\Omega_m^{0.6}\sigma_8$  
at $68 \%$ c.l..  
 
As we can see, a correlation appears from the present CMB data  
between these $2$ quantities: namely, increasing 
$\Omega_{cdm}h^2$ enhances the rms mass fluctuations. 
The independent constraint on  
$\Omega_m^{0.6}\sigma_8$ from clusters can be used to break the  
degeneracy. Using the {\it high} constraint, we obtain a higher 
value for the density in CDM (see Table 1) with  
$0.17 > \Omega_{cmd}h^2 > 0.11$ at $95 \%$ c.l.. 
Using the {\it low} value, the constraint becomes 
$0.15 > \Omega_{cmd}h^2 > 0.09$, again at $95 \%$ c.l..  
The $2$ results are consistent and  
 favour  CDM; however, when additional information such as SN-Ia and 2dF  
are included, the CMB tends to prefer the lower value. 

We further analyze this possible discrepancy in Figure 3 where we 
plot constraints in the $\Omega_m-\sigma_8$ plane
together with the
$68 \%$ c.l. results of \cite{pierpa} and of \cite{liddle}.
For values of $\Omega_m <0.35$, the {\it high}
$\sigma_8$ constraint is in slight disagreement with the
CMB+2dF result at more than $1 \sigma$.
 
On smaller (sub-galactic) scales, recent and strong constraints on 
the linear rms mass fluctuations $\sigma (M)$  
have been obtained through lensing measurements (~\cite{dalal}). 
In Figure $3$, we plot the predictions of the CDM models that  
are within the $95 \%$ CMB constraints in the 
$\sigma(R)-R$ plane together with a region indicative of the constraint 
obtained from lensing.  
A more careful comparison  and a better study of all the assumptions 
and possible systematics should  be done.  
However, here we wish to point out that, even in our simple 
analysis, the two contours overlap and the CDM paradigm seems consistent over  
a range of scales from $10^4 \rm Mpc \, h^{-1}$ to $10^{-1} \rm Mpc \, h^{-1}$. 
 
\begin{figure}[htb] 
\begin{center} 
\epsfxsize=7.2cm 
\epsfysize=6.2cm 
\epsffile{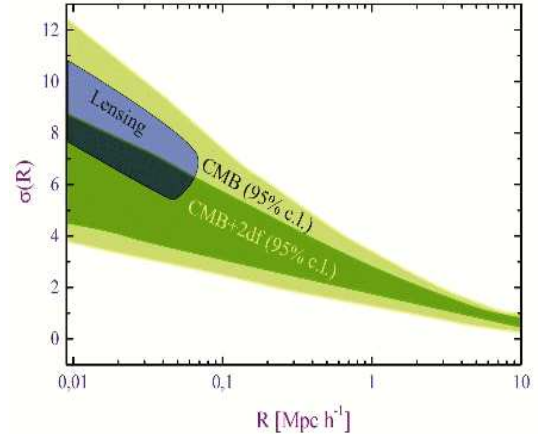} 
\end{center} 
\caption{The predictions of the CDM models that 
are within the $95 \%$ CMB constraints in the 
$\sigma(R)-R$ plane together with a region indicative of the constraint 
obtained from lensing.} 
\label{fig:like_mar2} 
\end{figure} 
\medskip 
 
\section{Discussion} 
 
\label{conclusion} 
 
In this paper we have provided strong constraints on the amount of 
non-baryonic dark matter in the universe by comparing adiabatic 
inflationary models with a set of cosmological observations. 
Combining CMB anisotropy measurements,  
high redshift supernovae observations, constraints  
on the Hubble parameter from the HST key project, and 
the matter power spectrum data from the 2dFGRS, we find  
$\Omega_{cdm}h^2=0.11_{-0.03}^{+0.04}$, $n_S = 0.93 \pm 0.08$,  
$\sigma_8^*=0.66\pm0.10$ at $95 \%$ confidence level. 
These results provide strong evidence for the presence of non-baryonic  
dark matter in  a way independent of the  
local cluster abundance data. 

If the dark matter is the neutralino, a knowledge of
$\Omega_{cdm}h^2$ can set strong bounds on the parameter space of the
simplest and most direct implementations of supersymmetry (see e.g.
\cite{ellis}, \cite{arnowitt}, \cite{rozkowski}).

The constraint on $\Omega_{cdm}h^2$ obtained here can be considered
 stable under the removal of some of the theoretical assumptions 
made in the analysis. 
Including quintessence, reionization and a large-scale CMB component
such as expected from gravity waves (see e.g. \cite{kmr},
\cite{efstathiougw}) or isocurvature CDM perturbations
(see e.g. \cite{trotta}) can relax the constraints but does not
change the conclusions. Models with $\Omega_{cdm}h^2$ as large as $0.2$
are disfavoured in all cases except under the exotic hypothesis of
an extra background of relativistic particles.

Depending on the external data we incorporate, 
the CMB constraints in the $\sigma_8\Omega_M^{0.6}-\Omega_{cdm}h^2$ 
plane can be $1-2 \sigma$ lower than most of the determinations
inferred from the local cluster X-ray 
temperature function (see e.g. \cite{pierpa}, \cite{borgani}, \cite{eke}) 
and cosmic shear data (\cite{vande}, \cite{maoli}, \cite{refregier}, 
\cite{bacon}), while  in better agreement
with the 'new' analyses of \cite{s8eljak} and \cite{liddle}.
However, we showed that  use of the {\it high} or
{\it low} priors on $\sigma_8$ has only a marginal effect on  the 
constraints on $\Omega_{cdm}h^2$.

With reference to previous  analyses, 
we are in agreement with the recent paper by 
Lahav et al. \cite{lahav}, although  
our value of $\sigma_8^*$  appears to be 
slightly lower. However we allow for variations in 
the spectral index $n_S$ while in \cite{lahav} this parameter has 
been set to $n_S=1$.
We also included the COBE data using the 
lognormal approximation as in \cite{bjk} and the 2dF data is taken 
from the independent analysis of \cite{thx}. 

The low values of  $\sigma_8$ and $n_S$ can possibly alleviate 
some of the problems of CDM on subgalactic scales.  
We will investigate this  in a forthcoming paper \cite{julien}. 
However, we also find that the models compatible with the CMB can  
satisfy the recent lensing constraints. This result  
suggests that a solution of the subgalactic CDM problem  
can most likely  be obtained by a refinement of  
the astrophysical processes involved in the numerical simulations, 
rather than by an {\it ad hoc} modification of the dark matter properties. 
However, if future cluster temperature or cosmic shear 
analyses were to converge towards a higher $\sigma_8$ value,
then this could lead to a possible discrepancy with the CMB+2dF result.
It will be the task of future experiments and analysis to verify this
interesting result.

\textit{Acknowledgments} 
 
The authors would like to thank Celine Boehm and Marco Peloso 
for useful discussions. AM acknowledges support from  PPARC.

\end{document}